\documentclass[twocolumn]{aastex631}

\shorttitle{Resonant chain of HD 110067}
\shortauthors{Lammers \& Winn}

\graphicspath{{./}{}}
\usepackage{amsmath}
\usepackage{comment}

\begin{document}

\title{The six-planet resonant chain of HD 110067}


\author[0000-0001-9985-0643]{Caleb Lammers}
\affiliation{Department of Astrophysical Sciences, Princeton University, 4 Ivy Lane, Princeton, NJ 08544, USA}

\author[0000-0002-4265-047X]{Joshua N.\ Winn}
\affiliation{Department of Astrophysical Sciences, Princeton University, 4 Ivy Lane, Princeton, NJ 08544, USA}

\begin{abstract}

HD 110067 is the brightest star known to have six transiting planets. Each adjacent pair of planets has a period ratio that is nearly equal to a ratio of small integers, suggesting the planets are in a chain of mean-motion resonances, but the limited time span of the available data has prevented firm conclusions. Here, we show that the requirement of long-term dynamical stability implies that all six planets are very likely to form a resonant chain. Dynamical simulations of nonresonant systems with initial conditions compatible with the available data almost always suffer an instability within $25$\,Myr (${\sim}\,0.3$\,\% of the system's age). Assuming the system is in resonance, we place upper limits on the planets' eccentricities and lower limits on the masses of the planets that have not yet been measured. We also predict the characteristics of transit timing variations and the values of the three-body libration centers.

\end{abstract}

\section{Introduction}
\label{sec:intro}

A ``resonant chain'' is a planetary system in which multiple adjacent planet pairs are locked in mean-motion resonances (MMRs). Although resonant chains are believed to be natural outcomes of slow migration in protoplanetary disks \citep[e.g.,][]{Henrard1982, Lee&Peale2002, Kley2005, Terquem&Papaloizou2007}, only ${\sim}\,1$\,\% of known exoplanetary systems are in a resonant chain configuration \citep{Fabrycky2014}. Because external perturbations tend to break resonances, resonant chains are believed to have avoided major disturbances since their formation and thereby provide a view of the immediate outcomes of the planet formation process.

Gliese 876 was the first known example of an exoplanetary resonant chain, with three planets --- two giant planets and a super-Earth-sized planet --- having periods nearly in the ratios 4:2:1 \citep{Rivera2010}. Later, Kepler-223 was found to host a four-planet resonant chain consisting of sub-Neptune-sized planets \citep{Mills2016}. Starting with the innermost planet and moving outward, the ratios of the periods of adjacent planets are nearly 4:3, 3:2, and 4:3. Nearly commensurable orbital periods do not guarantee, however, that a planetary system is trapped in a resonant configuration. Even planet pairs located within $1$\,\% of a first-order MMR (i.e., $|\Delta|\,{=}\,|\frac{j\,{-}\,k}{j}\frac{P_\mathrm{out}}{P_\mathrm{in}}\,{-}\,1|\,{<}\,0.01$, where $j$ and $k$ are the index and order, respectively, of a $j$:$j\,{-}\,k$ MMR) typically have circulating two-body resonant angles \citep{Hadden&Lithwick2017}.

\begin{table*}
\centering
\caption{Properties of the HD~110067 planetary system, as reported in \citet{Luque2023}. Planets c, e, and g have only $3\,\sigma$ upper limits on their masses, and all six eccentricities are unconstrained. Our $N$-body simulations were initialized based on the observed posteriors on the orbital parameters of the six planets (see Section~\ref{sec:nbody_sims} for more details).}
\begin{tabular}{ccccc}
 \hline
 Planet & Period [days] & Planet mass [$M_\oplus$] & Planet radius [$R_\oplus$] & Inclination [deg]\\
 \hline
 HD~110067 b & $9.113678\,{\pm}\,1\,{\times}\,10^{-5}$ & $5.69_{-1.82}^{+1.78}$ & $2.200\,{\pm}\,0.030$ & $89.061\,{\pm}\,0.099$\\
 HD~110067 c & $13.673694\,{\pm}\,2.4\,{\times}\,10^{-5}$ & ${<}\,6.3$ & $2.388\,{\pm}\,0.036$ & $89.687\,{\pm}\,0.163$\\
 HD~110067 d & $20.519617\,{\pm}\,4\,{\times}\,10^{-5}$ & $8.52_{-3.25}^{+3.31}$ & $2.852\,{\pm}\,0.039$ & $89.248\,{\pm}\,0.046$\\
 HD~110067 e & \tablenotemark{a}$30.793091\,{\pm}\,1.2\,{\times}\,10^{-5}$ & ${<}\,3.9$ & $1.940\,{\pm}\,0.040$ & $89.867\,{\pm}\,0.089$\\
 HD~110067 f & $41.05854\,{\pm}\,1\,{\times}\,10^{-4}$ & $5.04_{-1.94}^{+1.89}$ & $2.601\,{\pm}\,0.042$ & $89.673\,{\pm}\,0.046$\\
 HD~110067 g & \tablenotemark{a}$54.76992\,{\pm}\,2\,{\times}\,10^{-4}$ & ${<}\,8.4$ & $2.607\,{\pm}\,0.052$ & $89.729\,{\pm}\,0.073$\\
 \hline
\end{tabular}
\tablenotetext{a}{Periods determined based on two transits and the prior that the system is in a specific resonant chain configuration.}
\label{table:measurements}
\end{table*}

\citet{Mills2016} showed that the planets of Kepler-223 form a resonant chain based on the analysis of four years of observed transit timing variations (TTVs). Specifically, they demonstrated that the systems' three-body resonant angles (also termed ``Laplace angles'') are librating rather than circulating. TRAPPIST-1 is another well-characterized resonant chain system, which consists of seven approximately Earth-sized planets, the largest number of planets known in any resonant chain \citep{Gillon2017, Luger2017}. It is also possible for a resonant chain to consist of only a subset of a system's planets; for instance, the outer five planets of TOI-178 form a resonant chain, but the innermost planet is not part of the chain \citep{Leleu2021}.

HD~110067 is a K0 dwarf star with mass $M_\ast\,{\approx}\,0.8\,M_{\odot}$ and age $8\,{\pm}\,4$\,Gyr. With a $V$ magnitude of $8.4$, it is the brightest star known to have more than four transiting planets. \citet{Luque2023} discovered six transiting planets with sizes ranging from $1.9$ to $2.9$\,$R_\oplus$ and orbital periods ranging from $9.1$ to $55$\,days. Starting with the innermost planet, the period ratios between adjacent planetary orbits are nearly equal to 3:2, 3:2, 3:2, 4:3, and 4:3, suggesting that the system forms a resonant chain. If true, HD~110067 would be one of only three known systems with six or more planets trapped in a resonant chain, the other two being TRAPPIST-1 \citep{Gillon2017, Luger2017} and TOI-1136 \citep{Dai2023}. Under the assumption that the system forms a resonant chain, \citet{Luque2023} successfully scheduled follow-up observations of the transits of the two outer planets before their periods had been well established. However, because the number of observed transits is still relatively small, and because TTVs have not yet been detected, the case for a resonant chain in HD~110067 has not been established as securely as it was for TRAPPIST-1 and TOI-1136.

\begin{figure*}
\centering
\includegraphics[width=\textwidth]{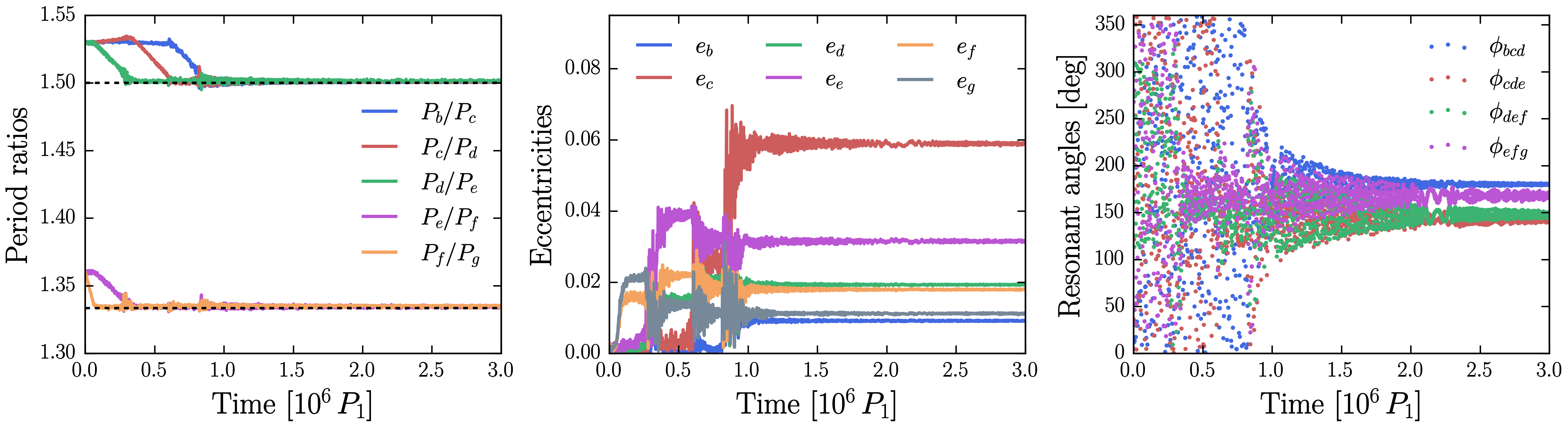}
\caption{Evolution of adjacent planet period ratios, orbital eccentricities, and three-body resonant angles during an example convergent migration simulation (in this case, $K\,{=}\,114$). Adjacent planet pairs are captured into two-body MMRs outside-in, beginning with planets f-g and ending with planets b-c. Each planet obtains a non-zero eccentricity through this process, and three-body resonant angles librate with small amplitudes (two-body resonant angles also librate). Before carrying out our long-term $N$-body simulations, we adiabatically remove the eccentricity and semi-major axis damping and provide a random kick to the planets' eccentricities.}
\label{fig:migration_example}
\end{figure*}

Due to the potential importance of the HD~110067 planetary system, dynamical characterization is warranted (see, e.g., \citealt{MacDonald2016, Mills2016, Tamayo2017, Siegel&Fabrycky2021, MacDonald2022, Quinn&MacDonald2023} for dynamical analyzes of other resonant chain systems). This Letter describes long-term $N$-body simulations that were undertaken to explore the dynamics of HD~110067. By assuming that the system is dynamically stable on timescales comparable to the age of the system, we wanted to know how confidently we may conclude that the system is a resonant chain. Since only three of the planets' masses have been measured (using the Doppler technique), and none of the eccentricities have been measured, we also wanted to know whether dynamical arguments can be used to place meaningful constraints on the masses and eccentricities.

\section{Simulation setups}
\label{sec:nbody_sims}

We have carried out $N$-body simulations of HD~110067-like systems initialized in two different ways: with parameters that are consistent with the available observations and no other constraints; and with parameters that emerge from a simple model of disk-driven migration into the observed configuration. Below, we describe the setup of our migration simulations.

The theory of resonant capture in two-planet systems is fairly well understood analytically \citep{Henrard1982, Borderies&Goldreich1984, Mustill&Wyatt2011, Batygin2015}, but the capture of higher-multiplicity systems into resonance is more complex and is typically modeled numerically. In this work, we adopt the migration model of \citet{Tamayo2017}, which allows us to generate many physically plausible resonant chains that are consistent with the observed properties of HD~110067. By initializing planets with random orbital orientations and sampling different choices of parameters governing the timescale and turbulence of migration, this procedure produces resonant chain systems that span a range of plausible eccentricities, deviations from exact resonant period ratios, and libration amplitudes. Similar approaches have been adopted in subsequent works \citep[e.g.,][]{MacDonald&Dawson2018, Siegel&Fabrycky2021, MacDonald2022}. Below, we describe our implementation of this migration model; more background on the method can be found in the works cited above.

At the outset of our simulations, the five outer planets were assigned periods that placed them $2$\,\% wide of the exact MMRs they are found near today (i.e., initial periods were chosen so that $\Delta\,{=}\,0.02$ for each adjacent planet pair). Each planet was initialized on a circular orbit, with an initial inclination and a mass drawn from independent normal distributions based on the posteriors reported in Table~\ref{table:measurements}. We set the stellar mass to $M_\ast\,{=}\,0.8\,M_{\odot}$ and drew the initial orbital orientations (i.e., true longitudes and longitudes of the ascending node) from a uniform distribution spanning [$0$, $2\pi$].

To capture the planets into a sequence of MMRs, we damped the semi-major axis of the outermost planet exponentially, with a damping timescale of $\tau_a\,{=}\,5\,{\times}\,10^6\,P_1\,{=}\,1.2\,{\times}\,10^5$\,years.\footnote{$\tau_a$ was chosen by manually increasing its value until the migration process was slow enough to successfully capture the six planets into a 3:2, 3:2, 3:2, 4:3, 4:3 resonance chain with a ${\sim}\,50$\,\% success rate.} The eccentricity of each planet was exponentially damped with a timescale $\tau_e\,{=}\,\tau_a/K$, where $K$ is a was drawn from a log-uniform distribution from $10$ to $10^3$. We integrated each system for one $\tau_a$ timescale and then adiabatically removed the eccentricity and semi-major axis damping over $5\,{\times}\,\tau_e$. To produce systems with a range of libration amplitudes (as expected due to turbulent perturbations during migration; e.g., \citealt{Rein&Papaloizou2009}), we then applied a parameterized kick to each of the planets' eccentricities. Specifically, we drew a random number $\eta$ from
a log-uniform distribution between $10^{-3}$ and $1$, and then either increased or decreased (with equal probability) each planet's eccentricity by a factor of $\eta$.

Finally, configurations that did not end up being captured into a 3:2, 3:2, 3:2, 4:3, 4:3 resonance chain were discarded. Configurations that successfully locked into the desired resonant chain were rescaled so that $P_1\,{=}\,9.11678$\,days. Fig.~\ref{fig:migration_example} shows an illustrative example of our migration procedure. Resonant capture in our simulations proceeds outside-in. First, planets f and g are captured into the 4:3 MMR, followed by planets e and f, and then the inner three planet pairs are sequentially captured into 3:2 MMRs. Over the course of this process, each planet acquires a new equilibrium eccentricity, which can be influenced by the capture of subsequent planets. After ${\sim}\,3\,{\times}\,10^6\,P_1$, the system is trapped in the resonant chain configuration, with librating two- and three-body resonant angles.

We then carried out long-term $N$-body integrations using the \texttt{WHFast} integrator \citep{Rein&Tamayo2015} from the open-source \texttt{REBOUND} package \citep{Rein2012} with a time step of $P_1/20$, where $P_1$ is the initial period of the planet b. Integrations were stopped when the Hill radii [$r_{H,\,i}\,{=}\,a_i (m_i/M_\ast)^{1/3}$] of two planets intersected, or after $10^9\,P_1\,{\approx}\,25$\,Myr, whichever came earlier. Damping forces were implemented using the \texttt{modify\_orbit\_forces} functionality from \texttt{REBOUNDx} \citep{REBOUNDx2020}.

\section{Long-term stability}
\label{sec:stability}

To simulate systems that are consistent with observations but are not necessarily in resonance, we integrated $350$ systems initialized by drawing parameters directly from the observational posteriors. In detail, we drew the initial periods from independent normal distributions with mean and standard deviation taken from Table~\ref{table:measurements}.\footnote{Since planets e and g have only been observed to transit twice, their periods are more uncertain than those of the other planets (see Fig.~S3 of \citealt{Luque2023}). Based on the period posteriors that were derived without assuming that planets e and g are in resonance, we adopted an uncertainty of $\Delta\,P\,{=}\,5$\,days on the periods of planets e and g. We also tried $\Delta\,P\,{=}\,1$\,day and found similar results.} Similarly, we drew the masses of the planets with radial-velocity (RV) measured masses (planets b, d, and e) from independent normal distributions, and we drew masses for the remaining three planets independently from [$0$, $m_\mathrm{max}$], where $m_\mathrm{max}$ is the $3\sigma$ RV mass upper limit from \citet{Luque2023}. Inclinations were also drawn independently from the posteriors (taking into account the discrete degeneracy in inclination inherent to transit observations) and eccentricities for each planet were drawn uniformly from [$0$, $0.05$], motivated by the eccentricity distribution that has been inferred for nonresonant multiplanet systems (\citealt{VanEylen&Albrecht2015, Hadden&Lithwick2017}; the initial eccentricities of our resonant chain systems span a similar range). Each longitude of pericenter and longitude of the ascending node was drawn randomly from [$0$, $2\pi$], and true longitudes were determined from the reported transit times. Mixed systems, in which the inner planets are resonant but the outer planets are not, were constructed by carrying out the migration process described above with a system of fewer planets, then adding the outer nonresonant planets to the system by drawing their properties from the observational posteriors.

\begin{figure}
\centering
\includegraphics[width=0.45\textwidth]{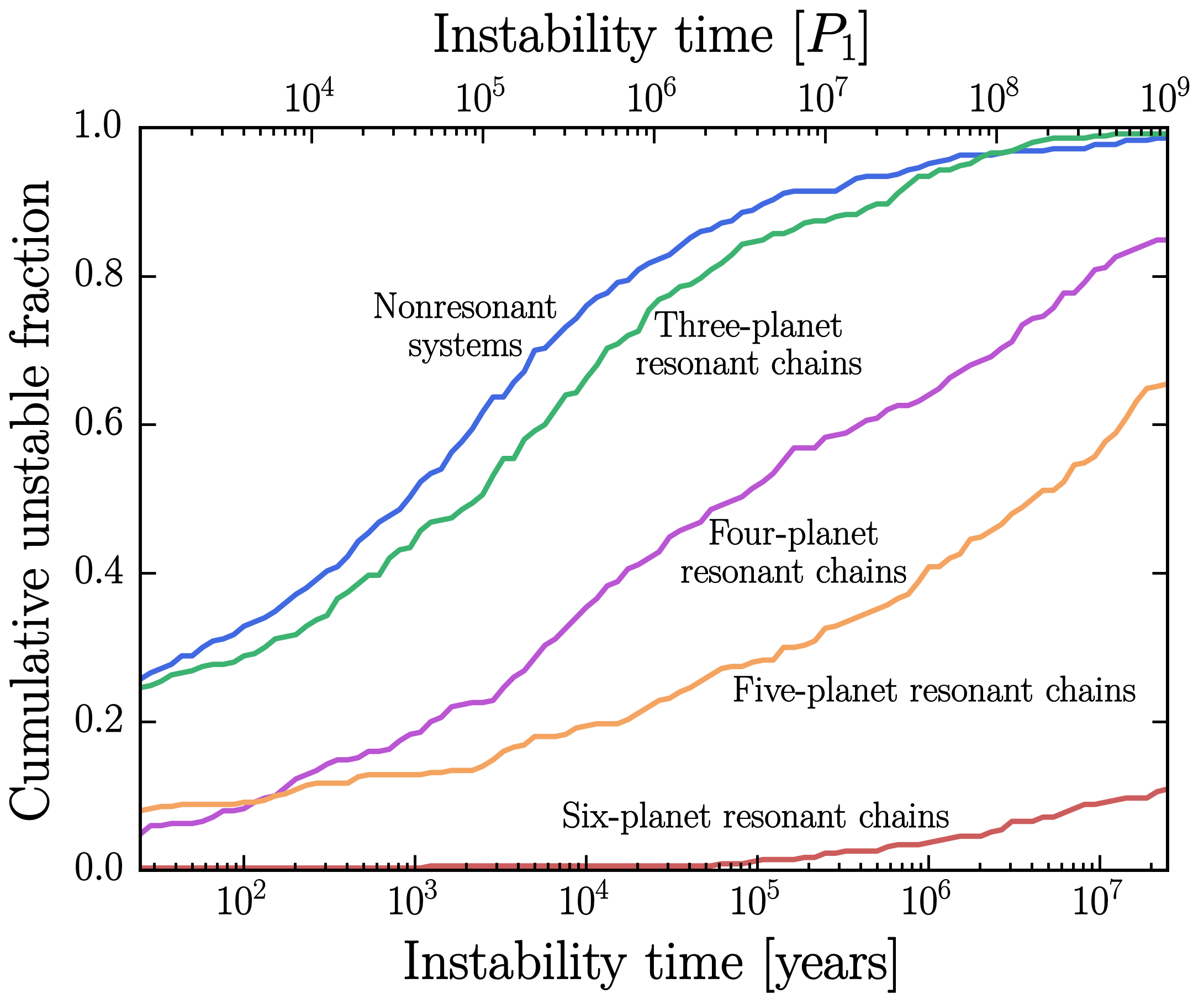}
\caption{Cumulative fraction of $350$ systems that destabilized within $25$\,Myr. $99$\,\% of the nonresonant systems with properties consistent with the observations were unstable, as opposed to $11$\,\% of systems initialized in a six-planet resonant chain (via a disk migration simulation). Systems in which only a subset of the planets are captured into a resonant chain were also mostly unstable.}
\label{fig:cumulative_t_insts}
\end{figure}

Figure~\ref{fig:cumulative_t_insts} shows the cumulative fraction of systems that proved to be unstable as a function of time. Systems with parameters drawn from the posteriors of \citet{Luque2023}, subject to no other constraints, were overwhelmingly unstable, with $345$ of the $350$ systems ($99$\,\%) destabilizing within $25$\,Myr (${\sim}\,0.3$\,\% of the system's age). In contrast, only $38$ of the $350$ systems ($11$\,\%) that underwent convergent migration into a six-planet resonance chain destabilized within $25$\,Myr. Systems in which the inner five planets formed a resonant chain but the outermost planet was decoupled from the resonant chain were also largely unstable, with $65$\,\% of such systems going unstable within $25$\,Myr. Decoupling more of the outermost planets from the resonant chain increased the fraction of systems that destabilized.

As a further experiment, we carried out $N$-body simulations in which a subset of the outer planets formed a resonant chain, and the inner planets were decoupled. This was achieved by carrying out migration simulations with a reduced number of planets, then adding the inner planets to the system by drawing their properties from the posteriors. With the innermost one, two, or three planets initialized directly from the observed posteriors, $41$\,\%, $58$\,\%, and $75$\,\% of systems destabilized within $25$\,Myr, respectively. We consider these results to be strong evidence that the six planets of HD~110067 form a six-planet resonant chain.

\section{Constraints on orbital architecture}
\label{sec:constraints}

Next, we explored the possibility of constraining the properties of the planets in HD~110067 by requiring that the six-planet resonant chain systems are long-term stable. The small sample size of the ensemble of simulations presented above was a barrier: there were only $38$ unstable systems out of the total of $350$ resonant chain systems, limiting our ability to find trends among the unstable systems. To expand the sample, we carried out $1050$ additional $N$-body simulations of six-planet resonant chain systems (of which $111$ destabilized), increasing the total number of resonant chain simulations to $1400$.

\begin{figure}
\centering
\includegraphics[width=0.45\textwidth]{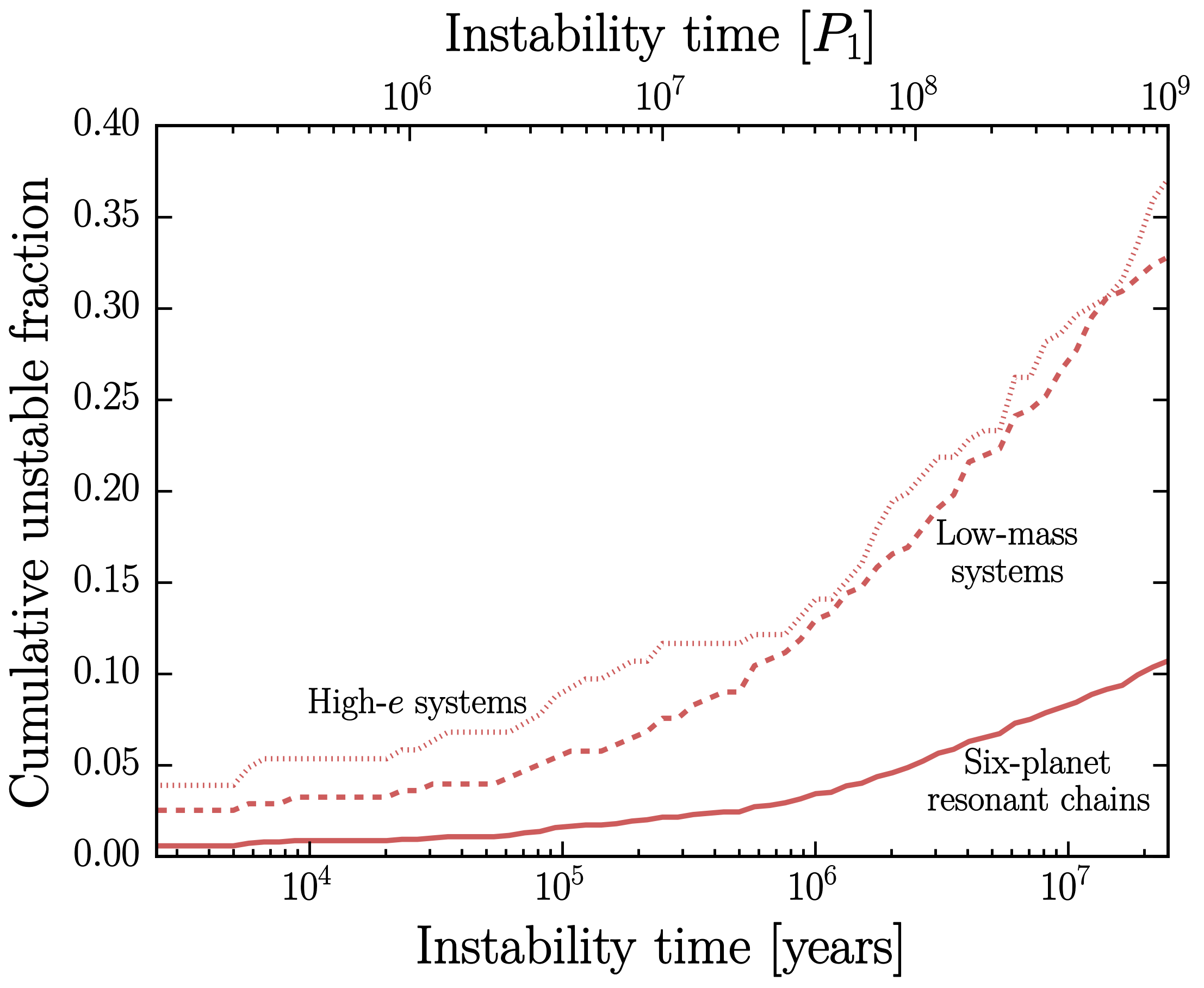}
\caption{The solid curve shows the cumulative unstable fraction of the $1400$ HD~110067-like resonant chain systems. The dashed and dotted curves, respectively, show the cumulative unstable fraction of the $278$ systems that contain a low-mass planet and the $206$ systems that contain a high-eccentricity orbit (i.e., systems with masses or eccentricities outside of the limits reported in Table~\ref{table:constraints}). Systems that contain a low-mass or high-eccentricity planet are significantly more likely to destabilize on long timescales, allowing us to place approximate lower limits on planet masses and upper limits on orbital eccentricities for the HD~110067 system.}
\label{fig:t_insts_constraints}
\end{figure}

First, we considered the influence of the planets' orbital eccentricities on the long-term stability of our synthetic HD~110067-like systems. Unsurprisingly, systems with larger eccentricities were more likely to destabilize, but the initial eccentricities of the six planets affected the stability of the systems to varying degrees. For our estimate of the upper limit on each planet's eccentricity $e_i$, we found the value of $e_\mathrm{max}$ for which more than $33$\,\% systems with $e_i\,{>}\,e_\mathrm{max}$ destabilized within $25$\,Myr. We report our upper limits on planet eccentricities in Table~\ref{table:constraints}. The weakest constraint is for planet c, probably because planet c is the only planet in the system in a 3:2 MMR (as opposed to 4:3) with both its inner and outer neighbors. The tightest constraint is on the eccentricity of planet g, probably because of the lack of the stabilizing influence of an outer resonant planet and the fairly tight (4:3) MMR with planet f. Figure~\ref{fig:t_insts_constraints} shows the cumulative unstable fraction of resonant systems that violate the eccentricity constraints given in Table~\ref{table:constraints}. The sharp upturn and rising trend in the unstable fraction over $25$\,Myr makes it unlikely that many of these configurations would survive for the full ${\sim}\,8$\,Gyr age of HD~110067.

Next, we turned to the influence of the planets' masses on long-term stability. Planets a, c, and e have RV-measured masses, but planets c, e, and g only have RV-based upper limits (see Table~\ref{table:constraints}). We did not find the systems for which $m_c$, $m_e$, or $m_g$ approached the RV mass upper limits to be especially unstable, indicating that the RV-based upper limits are more strict than the upper limits set by the requirement of long-term stability (although we did see an uptick in instability when $m_c$, $m_e$, or $m_g$ approached about $20\,M_\oplus$). On the other hand, systems in which a very {\it low} mass was assigned to planet c, e, or g were significantly more likely to destabilize on long timescales. This is probably because low-mass planets are more strongly perturbed by the gravitational interactions with the other planets in the system, making them more vulnerable to instability (see also Section~\ref{sec:discussion}).

Lower limits on the masses of planets c, e, and g, determined analogously to the eccentricity upper limits, are given in Table~\ref{table:constraints}. As was the case for the orbital eccentricities, the mass constraints are most stringent for planet g and least stringent for planet c. The cumulative fraction of unstable resonant chain systems that violate these mass lower limits increases rapidly at late times (see Fig.~\ref{fig:t_insts_constraints}). Systems that violate any of the eccentricity/mass constraints reported in Table~\ref{table:constraints} make up $86$\,\% of the unstable resonant chain simulations, whereas they account for $30$\,\% of all resonant chain systems in our sample. There is only moderate overlap between the high-eccentricity and low-mass samples; $67$ systems belong to both samples.

In addition to the properties of the planets, we attempted to constrain the values of $K$ and $\eta$, inspired by the observation that the unstable systems were initialized with preferentially low $K$ values and preferentially high $\eta$ values when compared with the stable systems. We found that requiring long-term stability leads to weak constraints on $K$ and $\eta$. Specifically, by requiring that fewer than $33$\,\% of systems should destabilize within $10^9\,P_1$, we found that $K\,{\gtrsim}\,13$ and $\eta\,{\lesssim}\,0.98$ (in other words, there are stable resonant chain configurations initialized with $K\,{\approx}\,10$\,--\,$1000$ and $\eta\,{\approx}\,0.001$\,--\,$1.00$). These relatively weak constraints prevent us from drawing strong conclusions about the convergent migration of HD 110067.

\section{Observational comparisons and future prospects}
\label{sec:TTVs}

\begin{table}
\centering
\caption{Constraints on the orbital eccentricities and planet masses of HD~110067 from the requirement that $<33$\,\% of six-planet resonant chains destabilize within $25$\,Myr. The masses of planets b, d, and f have been measured with RV data (see Table~\ref{fig:migration_example}).}
\begin{tabular}{ccc}
 \hline
 Planet & Eccentricity & Mass [$M_\oplus$] \\
 \hline
 HD~110067 b & ${\lesssim}\,0.07$ & $-$ \\
 HD~110067 c & ${\lesssim}\,0.15$ & ${\gtrsim}\,0.1$ \\
 HD~110067 d & ${\lesssim}\,0.08$ & $-$ \\
 HD~110067 e & ${\lesssim}\,0.08$ & ${\gtrsim}\,0.6$ \\
 HD~110067 f & ${\lesssim}\,0.08$ & $-$ \\
 HD~110067 g & ${\lesssim}\,0.04$ & ${\gtrsim}\,1.3$ \\
 \hline
\end{tabular}
\label{table:constraints}
\end{table}

\begin{figure}
\centering
\includegraphics[width=0.45\textwidth]{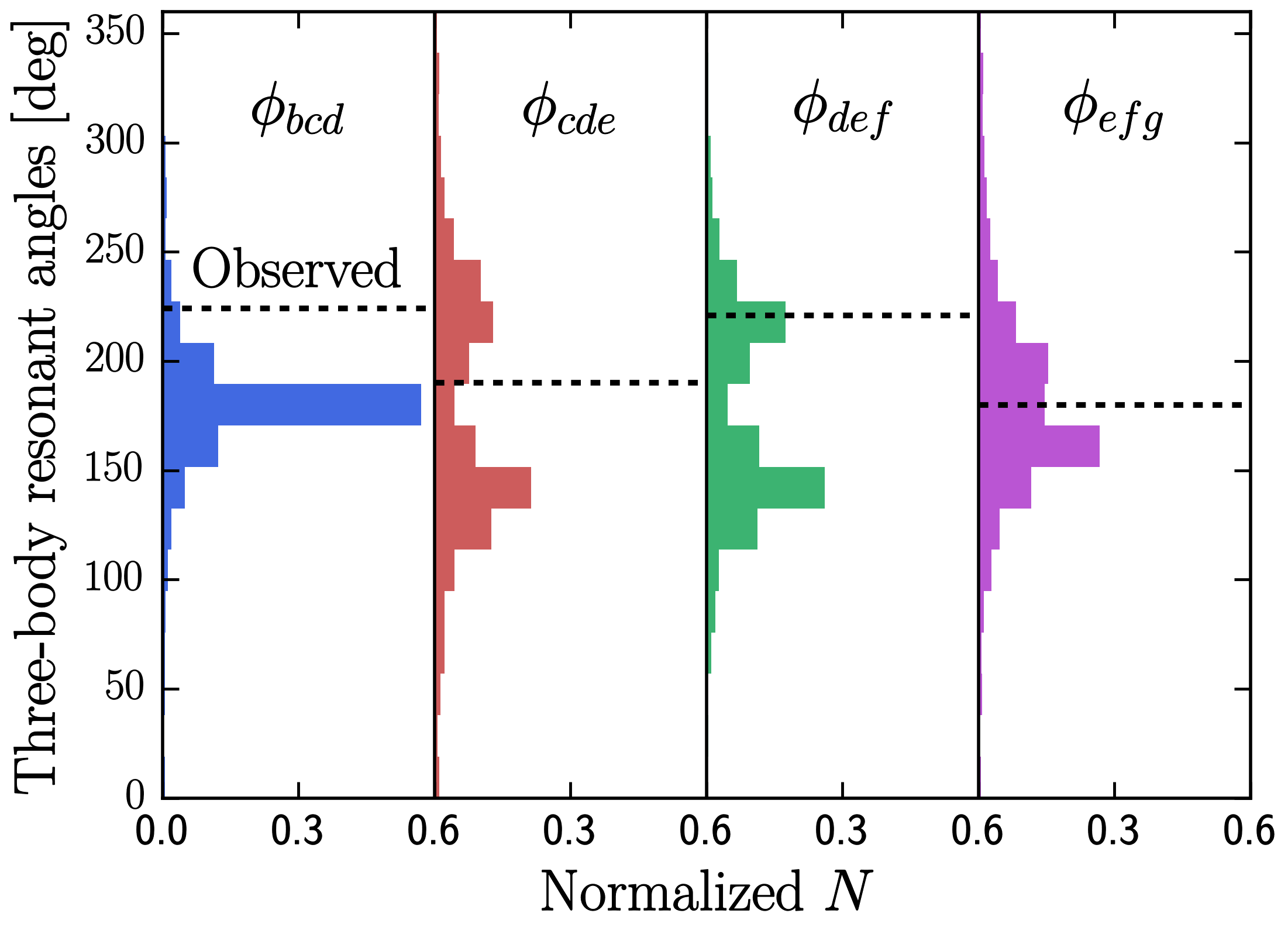}
\caption{Distributions of three-body resonant angle values in our $1251$ stable HD~110067-like resonant chain systems at the outset of the $N$-body simulations. Horizontal dashed lines mark the observed values, based on measured transit times. For all of the three-body resonant angles except $\phi_{bcd}$, the observed value falls comfortably within the range of resonant angles in our simulated resonant systems.}
\label{fig:3BR_angles}
\end{figure}

Unlike two-body resonant angles, which depend on the planets' longitudes of pericenter, three-body resonant angles depend only on the planets' mean longitudes and can therefore be determined directly from transit timings (e.g., $\phi_{bcd}\,{=}\,2\lambda_b\,{-}\,5\lambda_c\,{+}\,3\lambda_d$). Figure~\ref{fig:3BR_angles} shows the distribution of the three-body resonant angles across the stable resonant chain systems at the beginning of our $N$-body integrations. Because the resonant state of the system depends on how migration proceeds, different initial conditions can result in different libration centers and amplitudes. Assuming low orbital eccentricities, the observed transit times of the six planets (which have negligible uncertainty) can be used to determine the value of the mean longitudes, and therefore the three-body resonant angles, at one instant in time. The distributions of $\phi_{cde}$, $\phi_{def}$, and $\phi_{efg}$ are consistent with the observed values, although the observed value of $\phi_{bcd}$ is higher than $95$\,\% of the $\phi_{bcd}$ values in our simulated systems (a ``tension'' of $1.6\,\sigma$; see Section~\ref{sec:discussion}). Additional observations of HD~110067 are required to determine the true three-body libration centers and amplitudes.

Shifting our attention to the future, we expect that TTVs will be detected and will provide more precise constraints on the planets' masses and orbits. The TTVs of a planet pair that lie near a two-body $j$:$j\,{-}\,k$ MMR are expected to oscillate over the superperiod $P_s\,{=}\,\frac{1}{j}\frac{P_\mathrm{out}}{\Delta}$ \citep{Lithwick2012}. Due to the close proximity of planets to first-order MMRs in HD~110067 ($\Delta\,{\sim}\,10^{-4}$), superperiod oscillations would occur over long timescales (${\gtrsim}\,15$,$000$\,days), much longer than the current observational baseline. Resonant systems are expected to display TTVs over the two-body libration period $P_\ell\,{\propto}\,\left(\frac{m_1 + m_2}{M_\ast}\right)^{-2/3}$ \citep{Agol2005, Nesvorny&Vokrouhlicky2016}. Using Eq.~2 from \citet{Goldberg2022}, for HD~110067, we expect $P_\ell\,{\approx}\,2$,$000$\,--\,$6$,$000$\,days, depending on the planet pair.

\begin{figure*}
\centering
\includegraphics[width=\textwidth]{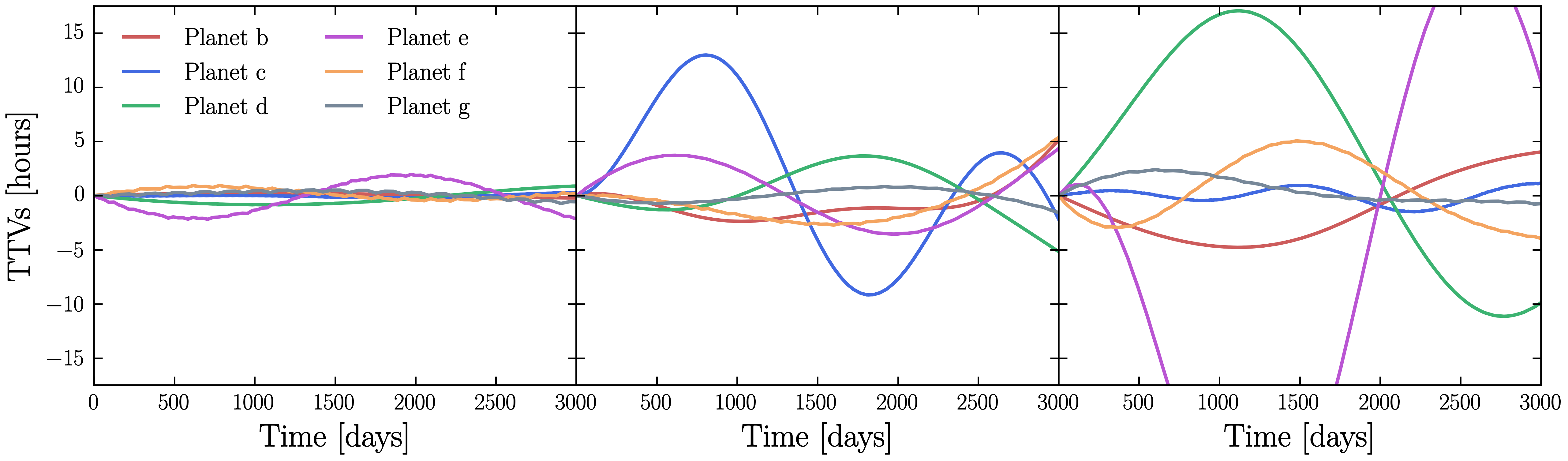}
\caption{Transit timing variations (TTVs) over $3000$\,days for three HD~110067-like resonant chain systems. The predicted TTVs typically evolve over timescales of thousands of days. The predicted amplitudes vary widely across our simulated systems, with a median peak-to-peak amplitude of $7_{-6.7}^{+15}$\,hr.}
\label{fig:TTV_example}
\end{figure*}

The amplitude of the TTV signals depends on the planets' proximity to resonance, masses, eccentricities, and orbital orientations. As a result, the predicted TTVs vary widely across our synthetic resonant chain systems. Commonly, the predicted TTVs show ${\sim}\,10$\,hr oscillations over thousands of days, in agreement with our analytical estimate of $P_\ell$. The median peak-to-peak amplitude and $1\,\sigma$ spread of the TTVs in our simulated systems is $7_{-6.7}^{+15}$\,hr. To illustrate the diversity of potential TTVs for HD~110067, Fig.~\ref{fig:TTV_example} shows the TTVs exhibited by three simulated resonant chain systems in our sample. Although there exist a few systems in our sample with negligible short-term TTVs, the typical TTV signal is large enough to be easily detectable with a several-year observational baseline.

\section{Discussion and conclusion}
\label{sec:discussion}

Thanks to the gigayear ages of typical observed exoplanet systems, investigations of long-term dynamical stability often help to characterize exoplanetary systems beyond what current observations can provide. Through such an investigation, we have concluded that HD~110067 is very likely to be trapped in a six-planet resonant chain. Assuming a resonant configuration, and requiring that the system be stable over timescales comparable to its age, we also derived constraints on orbital eccentricities and planet masses.

HD~110067 is notable among resonant chain systems in that when planetary parameters are drawn randomly from the distributions allowed by the available data, the resulting system is unstable at least $99$\,\% of the time. Previously, \citet{Hadden&Lithwick2017} noted that ${>}\,10$\,\% of $N$-body simulations of resonant chain systems Kepler-60 and Kepler-223 go unstable over $10^6\,P_1$. Similarly, \citet{Tamayo2017} found that ${\sim}\,40$\,\% systems initialized based on the observed posteriors of TRAPPIST-1 destabilized within $10^9\,P_1$, whereas systems initialized via disk migration simulations were largely stable over $10^9\,P_1$. HD~110067 is an even more extreme example of this trend. Part of the explanation is that HD~110067 was discovered recently and the current observational constraints are relatively weak. The dynamical fragility of the system is likely also related to the system's high multiplicity and compactness, with all adjacent planet period ratios $\frac{P_\mathrm{out}}{P_\mathrm{in}}\,{\lesssim}\,1.5$. The proximity of adjacent planets in the system to first-order MMRs ($\Delta\,{\approx}\,10^{-4}$) is also a factor, as instability times are known to shorten by several orders of magnitude in the vicinity of such MMRs \citep[e.g.,][]{Obertas2017, Petit2020, Lammers2024}.

Before carrying out our $N$-body simulations, we did not anticipate being able to place {\it lower} limits on any of the planet masses. Intuitively, one might think that since lowering a planet's mass weakens its mutual gravitational interactions, low planet masses promote long-term stability. However, this may not be true if there are other planets in the system with known masses, as is the case for HD~110067. Low-mass planets are more strongly perturbed by the other planets, potentially disturbing the resonant chain configuration. However, the exact dynamical mechanism for the instability is somewhat unclear. Theoretical works have suggested that resonant chain systems are destabilized by a resonance between a fast libration frequency and a slow difference between two synodic frequencies \citep{Pichierri&Morbidelli2020, Goldberg2022}. It is unclear why, in this dynamical picture, the inclusion of a low-mass planet would cause systems to destabilize; possibly, overstable librations play a role \citep{Goldreich&Schlichting2014}.

The observed value of the three-body angle $\phi_{bcd}\,{=}\,2\lambda_b\,{-}\,5\lambda_c\,{+}\,3\lambda_d\,{\approx}\,224^\circ$ is larger than $95$\,\% of the $\phi_{bcd}$ values in our simulated six-planet resonant chain systems, which cluster with fairly low amplitudes (${\sim}\,20^\circ$) about the $180^\circ$ equilibrium. Although the statistical significance of this ``tension'' is modest, we wondered about the possible reasons, given that previous works were able to reproduce the observed three-body resonant angles of other resonant chains using similar methods \citep[e.g.,][]{Mills2016, Tamayo2017}. Through experimentation, we found that the discrepancy is not significantly reduced by modifying the range of $K$ and $\eta$ values in our migration simulations. The large observed value of $\phi_{bcd}$ can be explained if $\phi_{bcd}$ is circulating, rather than librating. However, if the planets b, c, and d are not trapped in a three-body resonance, our $N$-body simulations indicate that the system would likely be unstable. With a shortest period of $P\,{\approx}\,9$\,days, it is also unlikely that tides from the host star are affecting the dynamics of the system.

Could the discrepancy be caused by an undetected inner planet in the HD~110067 system? To explore this possibility, we carried out migration simulations with a $1$\,--\,$10\,M_\oplus$ innermost planet in a first-/second-order MMR with planet b. We found that including an additional inner planet in the system does widen the distribution of $\phi_{bcd}$ values, and relaxes the tension with the observed value from $1.6\,\sigma$ to about $1.0\,\sigma$. Obviously, this is not yet a compelling hypothesis, but it should be kept in mind as additional data are collected. Given the system's observational accessibility and rich dynamical structure, HD~110067 promises to be a popular system for observers and theorists alike.

\section{Acknowledgments} 
\label{sec:acknowledgments}

We thank the anonymous referee for valuable comments. We also thank Jared Siegel and Max Goldberg for useful discussions. The simulations presented in this Letter were performed on computational resources managed and supported by Princeton Research Computing, a consortium of groups including the Princeton Institute for Computational Science and Engineering (PICSciE) and the Office of Information Technology's High Performance Computing Center and Visualization Laboratory at Princeton University.


\bibliography{refs}{}
\bibliographystyle{aasjournal}

\end{document}